\documentclass[prl,superscriptaddress,twocolumn]{revtex4-1}
\usepackage{tikz}
\usepackage{graphicx}
\usepackage{amsmath}
\usepackage{gensymb}
\usepackage[margin=1cm]{geometry}
\usepackage{setspace}
\usepackage{xcolor}
\usepackage{cancel}
\usepackage{soul}
\usepackage{wrapfig}
\usepackage{ulem}

\newcommand{\reftextit}[1]{}
\newcommand{\bra}[1]{\left< #1 \right|}
\newcommand{\ket}[1]{\left| #1 \right>}

\newcommand{\bk}{\mathbf{k}}
\newcommand{\bK}{\mathbf{K}}
\newcommand{\bq}{\mathbf{q}}
\newcommand{\bQ}{\mathbf{Q}}
\newcommand{\bG}{\mathbf{G}}
\newcommand{\bR}{\mathbf{R}}

\begin{document}
\title{Twist Angle Dependent Interlayer Exciton Lifetimes in van der Waals Heterostructures}

\author{Junho Choi}
\thanks{J.C. and M.F. contributed equally to this work.}
\affiliation{Department of Physics and Center for Complex Quantum Systems, The University of Texas at Austin, Austin, TX 78712, USA.}
\author{Matthias Florian}
\thanks{J.C. and M.F. contributed equally to this work.}
\affiliation{Institute for Theoretical Physics, University of Bremen, 28334 Bremen, Germany.}
\author{Alexander Steinhoff}
\affiliation{Institute for Theoretical Physics, University of Bremen, 28334 Bremen, Germany.}
\author{Daniel Erben}
\affiliation{Institute for Theoretical Physics, University of Bremen, 28334 Bremen, Germany.}
\author{Kha Tran}
\affiliation{Department of Physics and Center for Complex Quantum Systems, The University of Texas at Austin, Austin, TX 78712, USA.}
\author{Dong Seob Kim}
\affiliation{Department of Physics and Center for Complex Quantum Systems, The University of Texas at Austin, Austin, TX 78712, USA.}
\author{Liuyang Sun}
\affiliation{Department of Physics and Center for Complex Quantum Systems, The University of Texas at Austin, Austin, TX 78712, USA.}
\author{Jiamin Quan}
\affiliation{Department of Physics and Center for Complex Quantum Systems, The University of Texas at Austin, Austin, TX 78712, USA.}
\author{Robert Claassen}
\affiliation{Department of Physics and Center for Complex Quantum Systems, The University of Texas at Austin, Austin, TX 78712, USA.}
\author{Somak Majumder}
\affiliation{Materials Physics \& Applications Division: Center for Integrated Nanotechnologies, Los Alamos National Laboratory, Los Alamos, New Mexico 87545, USA.}
\author{Jennifer A. Hollingsworth}
\affiliation{Materials Physics \& Applications Division: Center for Integrated Nanotechnologies, Los Alamos National Laboratory, Los Alamos, New Mexico 87545, USA.}
\author{Takashi Taniguchi}
\affiliation{International Center for Materials Nanoarchitectonics, National Institute for Materials Science,\\ 1-1 Namiki, Tsukuba, Ibaraki
305-0044, Japan.}
\author{Kenji Watanabe}
\affiliation{Research Center for Functional Materials, National Institute for Materials Science,\\ 1-1 Namiki, Tsukuba, Ibaraki
305-0044, Japan.}
\author{Keiji Ueno}
\affiliation{Department of Chemistry, Graduate School of Science and Engineering, Saitama University, Saitama, 338-8570, Japan.}
\author{Akshay Singh}
\affiliation{Department of Physics, Indian Institute of Science, Bengaluru, Karnataka 560012, India.}
\author{Galan Moody}
\affiliation{Department of Electrical and Computer Engineering, University of California Santa Barbara, Santa Barbara, CA 93106, USA.}
\author{Frank Jahnke}
\affiliation{Institute for Theoretical Physics, University of Bremen, 28334 Bremen, Germany.}
\author{Xiaoqin Li}
\thanks{Corresponding author}
\email{elaineli@physics.utexas.edu}
\affiliation{Department of Physics and Center for Complex Quantum Systems, The University of Texas at Austin, Austin, TX 78712, USA.}

\begin{abstract}
 In van der Waals (vdW) heterostructures formed by stacking two monolayers of transition metal dichalcogenides, multiple exciton resonances with highly tunable properties are formed and subject to both vertical and lateral confinement. We investigate how a unique control knob, the twist angle between the two monolayers, can be used to control the exciton dynamics. We observe that the interlayer exciton lifetimes in $\text{MoSe}_{\text{2}}$/$\text{WSe}_{\text{2}}$ twisted bilayers (TBLs) change by one order of magnitude when the twist angle is varied from 1$^\circ$ to 3.5$^\circ$. Using a low-energy continuum model, we theoretically separate two leading mechanisms that influence interlayer exciton radiative lifetimes. The shift to indirect transitions in the momentum space with an increasing twist angle and the energy modulation from the moir\'e potential both have a significant impact on interlayer exciton lifetimes. We further predict distinct temperature dependence of interlayer exciton lifetimes in TBLs with different twist angles, which is partially validated by experiments. While many recent studies have highlighted how the twist angle in a vdW TBL can be used to engineer the ground states and quantum phases due to many-body interaction, our studies explore its role in controlling the dynamics of optically excited states, thus, expanding the conceptual applications of ``twistronics''.
\end{abstract}

\maketitle
Van der Waals (vdW) heterostructures offer a unique material platform with rich electronic and optical properties highly tunable via a wide selection of layer composition, strain, electric gating, and doping \cite{Novoselov:2016}. Because the lattice matching restriction is lifted at the interface, the twist angle between two monolayers (MLs) has emerged as a unique control knob to engineer the moir\'e superlattice formed by periodic variations of atomic alignment between the two layers \cite{Liu:2016, Quan:2020}. Following exciting discoveries in graphene twisted bilayers (TBLs) \cite{Cao:2018:1,Cao:2018:2, Kim:2017}, transition metal dichalcogenide (TMD) TBLs have also been found to exhibit rich correlated electronic phases \cite{Tang:2020,Regan:2020,Shimazaki:2020,Wang:2020}. In a different context, the lateral confinement introduced by the moir\'e potential may be used to realize a regular array of quantum emitters, a long-standing goal in the field of solid-state quantum information technology \cite{Yu:2017, Wu:2018}.

\begin{figure}
\centering
\includegraphics[width=0.48\textwidth]{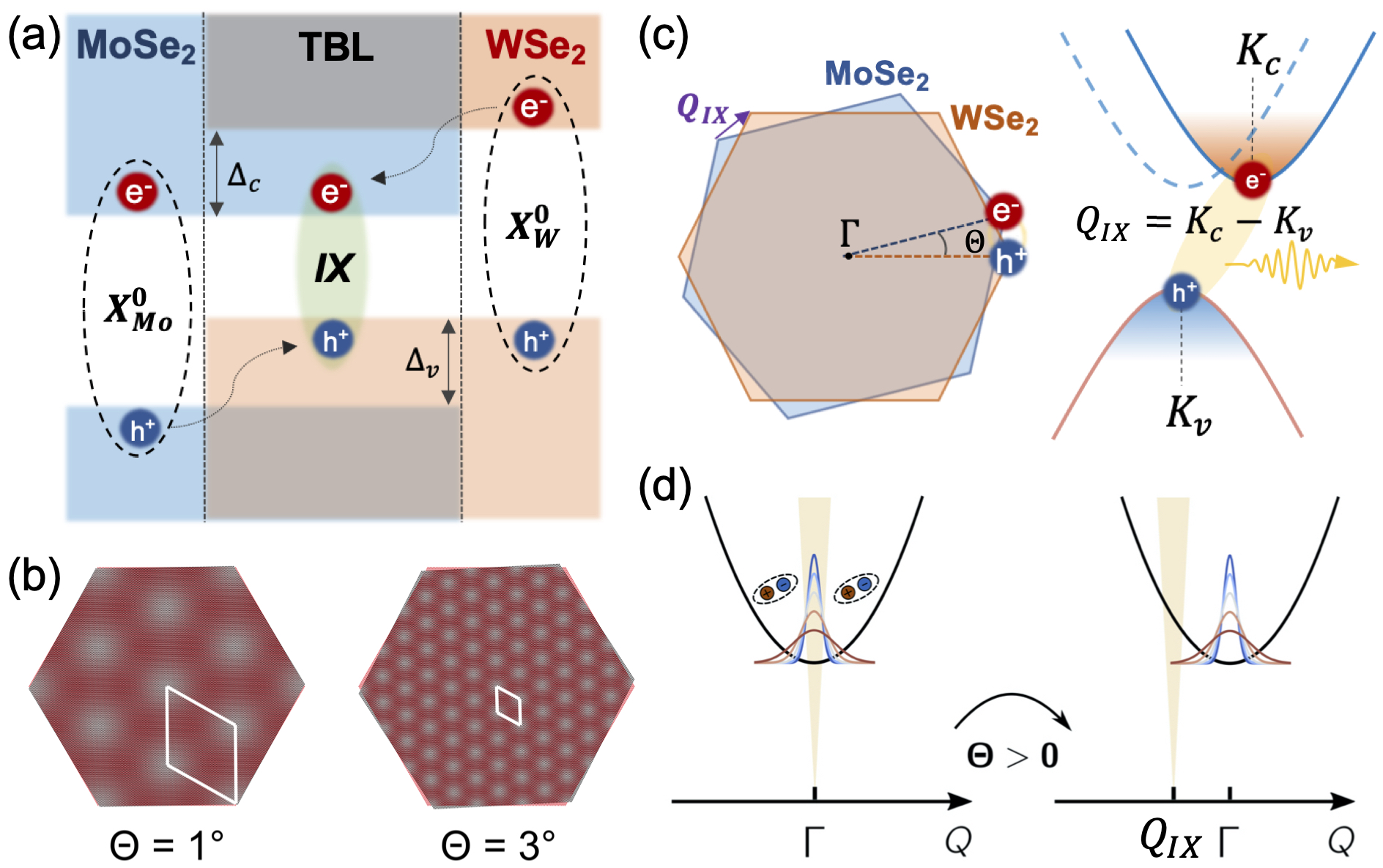}
\caption{\label{fig1} Conceptual description of IXs in a TBL. (a) Schematic of type-II band alignment with band offsets ($\Delta_c$ and $\Delta_v$) and interlayer charge transfer lead to the formation of IXs. (b) Real-space representation of the moir\'e superlattice for twist angles of 1$^\circ$ and 3$^\circ$ with a moir\'e supercell marked in white.  (c) Schematic of the Brillouin zone of $\text{MoSe}_{\text{2}}$ ML (blue) and $\text{WSe}_{\text{2}}$ ML (orange) twisted in the momentum space (left panel). $Q_{IX}=K_c - K_v$ denotes a finite momentum mismatch between the two layers and corresponds to the center of mass momentum of IXs. This twist leads to an indirect transition (right panel). (d) Thermal distribution of IXs with a near-zero (left) and a finite (right) twist angle plotted in the exciton's reference frame. Blue (red) curve illustrates the thermal distribution of IXs at a low (high) temperature. Yellow shaded region illustrates the light cone.}
\end{figure}

A type-II band alignment is typically found in a TMD TBL (Fig.~1a), leading to the formation of both intralayer and interlayer excitons \cite{Kang:2013, Rivera:2015, Tran:2019, Seyler:2019, Alexeev:2019}. We focus on interlayer excitons (IXs) because they are most likely to experience a deep moir\'e potential confinement. Because of the similar lattice constants of $\text{MoSe}_{\text{2}}$ and $\text{WSe}_{\text{2}}$ MLs, the size of the moir\'e supercell is determined by the twist angle (Fig.~\ref{fig1}b). The twist angle in real space translates into a rotation in the momentum space, causing the relative rotation of Brillouin zones associated with each monolayer (Fig.~\ref{fig1}c). Consequently, valleys in each layer are shifted from each other, changing a direct optical transition near the K-valley to an indirect transition, introducing longer IX lifetimes \cite{Rivera:2015, Yu:2015}. We illustrate the exciton center of mass wavefunction in its own reference frame (i.e., the $\Gamma$ point) in the excitation picture of Fig.~\ref{fig1}d. In TBLs with small twist angles ($\Theta< 5^\circ$), the thermally broadened IX distribution is sufficient to satisfy the momentum conservation requirement of an indirect transition, allowing the radiative decay process. A wide range of IX lifetimes in TMD TBLs have been reported in time-resolved photoluminescence (TRPL) measurements previously \citep{Tran:2019, Rivera:2015, Rivera:2016, Miller:2017, Nagler:2017, Jauregui:2019, Wang:2019, Choi:2020}. However, no prior studies have explained the origin of such variations.

We perform TRPL measurements on a series of $\text{MoSe}_{\text{2}}$/$\text{WSe}_{\text{2}}$ TBLs with accurately controlled twist angles. And we observe that the IX lifetime changes by one order of magnitude when the twist angle is changed from 1$^\circ$ to 3.5$^\circ$. Theoretically, we examine two mechanisms that influence the IX radiative lifetimes. We find that both the shift to indirect optical transition with an increasing twist angle and the moir\'e potential have a significant effect on the lifetime. Our theory further predicts that IX lifetimes exhibit different temperature dependence in TBLs with different twist angles. This prediction is partially validated by the experimental observations.

\begin{figure}
\centering
\includegraphics[width=0.48\textwidth]{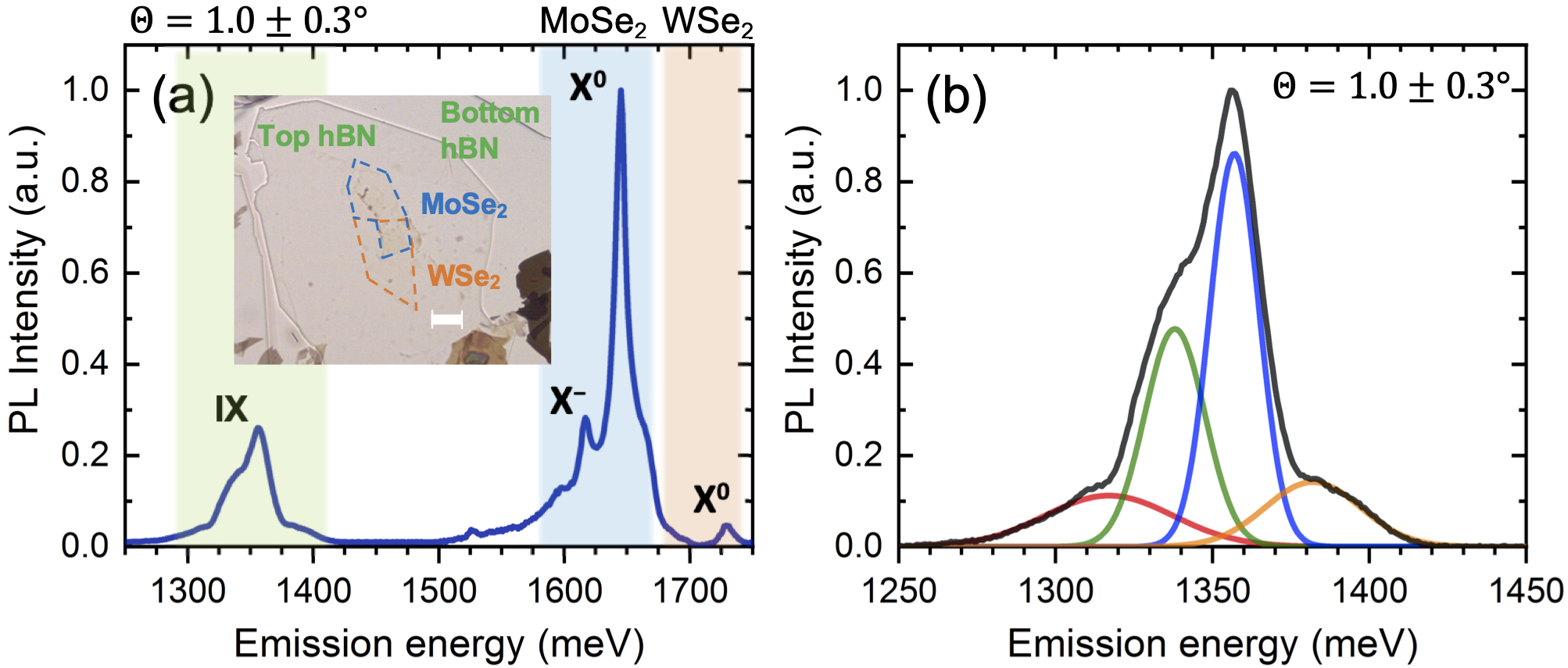}
\caption{\label{fig2} (a) Low-temperature PL spectrum from a $\text{MoSe}_{\text{2}}$/$\text{WSe}_{\text{2}}$ TBL with $\Theta = 1.0 \pm 0.3 ^\circ$ twist angle, featuring both intralayer excitons (neutral exciton: $X^{0}$, trion: $X^{-}$) and IXs. Inset shows the optical image of the hBN-encapsulated TBL. $\text{MoSe}_{\text{2}}$ and $\text{WSe}_{\text{2}}$ ML regions are indicated inside the blue and orange dashed lines, respectively. Scale bar: 20 $\mu m$ (b) Gaussian fitting to multiple IXs. Black solid line is measured data.}
\end{figure}

The hBN encapsulated TBLs are prepared using a standard exfoliation and stacking procedure described in the supplementary materials (SM) \cite{Andres:2014}. The twist angle is estimated from high-resolution optical microscopy images, and the stacking type of the TBLs is determined by second-harmonic generation (SHG) (see SM for details) \cite{Kumar:2013,Malard:2013,Li:2013,Hsu:2014}. All optical experiments are performed using an averaged excitation power of 1 $\mu$W and a focused laser spot size of $\sim$ 1 $\mu$m in diameter. Our far-field experiment probes hundreds of moir\'e supercells simultaneously. The data represent statistical averages of an ensemble of excitons. We present the low-temperature PL spectrum from the TBL with $\Theta = 1.0 \pm 0.3 ^\circ$ twist angle in Fig.~\ref{fig2}a, in which both IXs and intra-layer exciton resonances with large binding energy of a few hundred meV are observed \cite{Wilson:2017, Chiu:2014, Chernikov:2014, Ugeda:2014, Tran:2019, Seyler:2019, Rivera:2015}. We fit multiple IX resonances in the PL spectrum by Gaussian functions as shown in Fig.~\ref{fig2}b. We present the PL spectra for the other two samples with different twist angles, and the analysis of the multiple IX resonances with consistent dielectric environments in the SM \cite{Gerber:2018,Raja:2019}. Different possible interpretations for these resonances have been proposed, including phonon-mediated states, defect-bound states, and spin triplet states \cite{Danovich:2018, Zhang:2019, Ciarrocchi:2019, Li:2020}. While we cannot rule out these alternative explanations definitively, we find that the interpretation of quantized exciton resonances confined within the moir\'{e} potential captures the essential features of our data as we further explain below \cite{Wu:2018, Tran:2019, Brem:2020}. 

\begin{figure}
\centering
\includegraphics[width=0.48\textwidth]{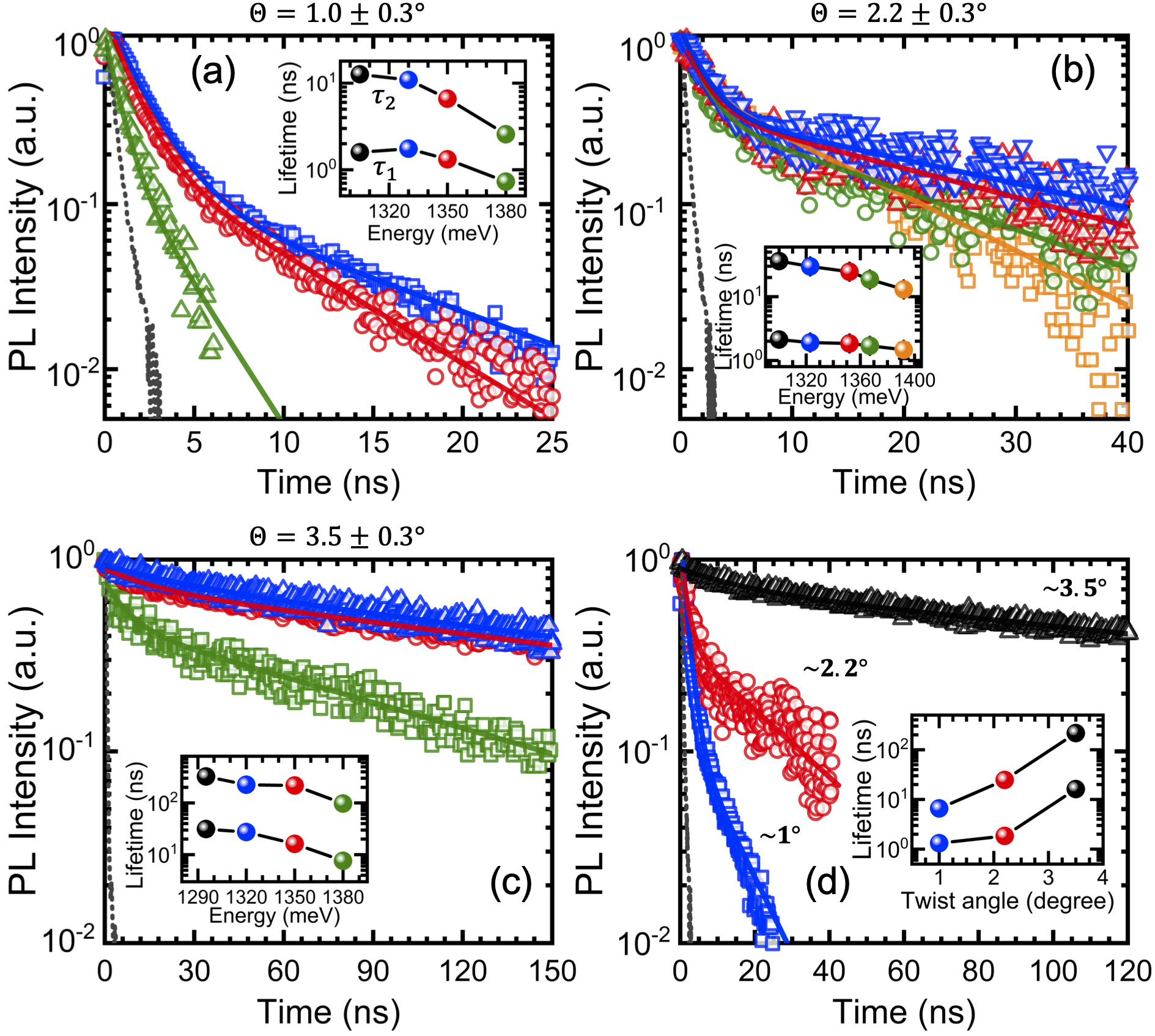}
\caption{\label{fig3}Twist angle dependent lifetimes of all IX resonances measured from (a) $\Theta = 1.0 \pm 0.3 ^\circ$, (b) $2.2 \pm 0.3 ^\circ$, and (c) $3.5 \pm 0.3 ^\circ$ TBLs, respectively. Inset shows the energy dependence of the fast ($\tau_{1}$) and slow ($\tau_{2}$) decay components of IX lifetimes. Solid and black dotted lines represent fitting with a biexponential decay and instrumental response functions, respectively. (d) Twist angle dependence of IX lifetimes measured near 1345 - 1355 meV in the TBLs. Inset summarizes the extracted fast and slow decay components in the three TBLs.}
\end{figure}

The recombination dynamics of IXs in three TBLs are spectrally and temporally resolved at low temperatures as shown in Fig.~\ref{fig3}a-c. The TRPL signals are fitted with a biexponential decay function. Such a biexponential decay indicates that a simple two-level model is not sufficient to explain the dynamics of IXs. This biexponential decay may originate from scattering between a bright and dark exciton \cite{Moody:2016, Palummo:2015, Zhang:2015}. Using a rate equation analysis for a three-level system (see SM for details, which includes Refs. \cite{Dalgarno:2005,Chen:2008, Hanson:2015,Orfield:2018}), we attribute the fast (slow) decay component to the lifetime of the bright (dark) exciton assuming the scattering rate between the bright and dark state is slower than the bright exciton lifetime. In all samples, we observe shorter IX lifetimes as the energy of the resonance increases as shown in the inset to each panel in Fig.~\ref{fig3}a-c. This systematic change with energy is consistent with the interpretation that the higher energy excitons are excited states in the moir\'{e} potential with additional relaxation channels \citep{Tran:2019}. The reduced exciton lifetimes associated with the excited states are often found in spatially localized excitons in other materials \cite{Steinhoff:2013}.

Most remarkably, the IX lifetime changes drastically as a function of the twist angle, as shown in Fig.~\ref{fig3}d. Both the fast and slow components of the IX lifetime increase nearly by one order of magnitude when the twist angle increases from $\Theta$ = 1$^\circ$ to 3.5$^\circ$. For example, the fast (slow) decay time measured over the range of 1345 - 1355 meV increases from 1.3 (6.6) to 16.2 (217.5) ns. We focus on understanding this twist angle dependence of IX lifetimes in TBLs.

To understand this drastic twist angle dependence, we calculate the radiative recombination rate of IXs in the TBLs (see the SM for details, which includes Refs. \cite{liu_three-band_2013,Wang:2019,florian_dielectric_2018,asriyan_optical_2019,semina_excitons_2019,kylanpaa_binding_2015,geick_normal_1966,rytova1967the8248,keldysh_coulomb_1979}). We note that the lifetime measured is in fact the total lifetime, determined by both radiative and non-radiative processes. It is impractical to calculate all non-radiative processes, which may depend on many extrinsic properties. Thus, our calculations can only be compared to ideal emitters with unity quantum yield. Nevertheless, these calculations provide important understanding of IX dynamics controlled by the twist angle in a TBL and guide the search for more efficient emitters in a moir\'e crystal. Our calculations start with Fermi's golden rule~\cite{esser_photoluminescence_2000}

\begin{eqnarray}
\tau^{-1}_{\text{IX}} = \frac{2\pi}{\hbar} \sum\limits_{if}| \bra{f} H_{\text{rad}} \ket{i} |^2 \delta(\varepsilon_i - \varepsilon_f) N_i\,
\label{eq:IX_fermisgoldenrule0}
\end{eqnarray}
where the optical matrix element describes transitions between the initial IX state $\ket{i}$ and the final photon state $\ket{f} = b^\dagger_{\bq, \sigma}\ket{0}$. Here,  $b^\dagger_{\bq, \sigma}$ creates a photon with wave vector $\mathbf q$ and polarization $\sigma$ with respect to the vacuum state $\ket{0}$, and $N_i$ is the occupation of the initial IX states. To describe the moir\'{e} exciton states we use a low energy continuum model that we derive along the lines of Ref.~\cite{Yu:2015, Wu:2017, Yu:2017, Wu:2018}. As the basis of describing the hybridized IXs later, we have first written the wave function of IXs in the absence of interlayer coupling as:
\begin{eqnarray}
\ket{\bQ} &=& \frac{1}{\sqrt{\mathcal{A}}}\sum\limits_{\bk} \phi_X(\bk) {a}^{\dagger,c}_{\bK_c+\bk + \frac{m_e}{M}\bQ} {a}^v_{\bK_v+\bk-\frac{m_h}{M}\bQ} \ket{0} \label{eq:X_wfct}
\end{eqnarray}
with eigenenergies $E(\bQ)= \hbar^2|\bQ|^2/(2M) + E_{\text{gap}} - E_{B}$ that are characterized by the center-of-mass momentum $\bQ$. Here, ${a}^{\dagger,c}_{\bK_c + \bq} ({a}^{v}_{\bK_v + \bq})$ creates (annihilates) an electron in the K$_c$ (K$_v$)-valley of the conduction (valence) band, $M=m_e+m_h$ is the total exciton mass, and $m_e$ ($m_h$) is the electron (hole) effective mass. $\mathcal{A}$ is the crystal area. The electron-hole relative-motion wave function in momentum space $\phi_X(\bk)$ is determined by the Wannier equation. Its solution determines the IX energy $E_{\text{gap}} - E_{B}$, where $E_{B}$ is the IX binding energy and $E_{\text{gap}}$ is the band gap.

An interlayer twist in real space generates a relative shift in momentum space ($\bK_c - \bK_v$, see Fig.~1c) as well as a spatial modulation of the exciton energy. In all three TBLs, the moir\'{e} periodicity can be assumed to be large compared to the IX Bohr radius. Thus, we neglect the variation of the binding energy in the moir\'{e} pattern and use a local approximation of the IX moir\'{e} potential $V^{\textrm{M}}(\bR)$ according to Ref.~\cite{Wu:2018}. In this framework, the IX Hamiltonian reads
\begin{eqnarray}
H &=& -\frac{\hbar^2}{2 M}\Delta_{\bR} + V^{\textrm{M}}(\bR)\,,
\label{eq:moire_hamiltonian}
\end{eqnarray}
where  $-\frac{\hbar^2}{2 M}\Delta_{\bR}$ is the exciton center-of-mass kinetic energy and can be diagonalized using a plane-wave expansion with eigenenergies $\varepsilon_{\bQ,\lambda}$ and eigenstates
\begin{eqnarray}
\ket{ \bQ, \lambda } = \sum\limits_{\bG_{\text{M}}} c^{\lambda }_{\bQ-\bG_{\text{M}}} \ket{\bQ - \bG_{\text{M}}}\,.
\label{eq:plane_wave}
\end{eqnarray}
The moir\'{e} reciprocal lattice vectors $\bG_{\text{M}}$ are derived as differences of top and bottom layer reciprocal lattice vectors. Therefore, the size of the moir\'{e} Brillouin zone (MBZ) scales with the interlayer twist angle. We represent its center and boundary by $\gamma$ and $\kappa$, respectively. We use the above plane-wave expansion, and solve the optical matrix element between the initial moir\'{e} exciton state $\ket{\bQ,\lambda}$ and the final photon state $\ket{\bq,\sigma}$. Moiré excitons can only recombine if the momentum conservation law $\bQ - \bq_\parallel= \bK_c - \bK_v := \bQ_{IX}$ is fulfilled where $\bq_\parallel$ represents the in-plane component of the photon wavevector. We assume that the twist angle is sufficiently large so that typical photon momenta are still small compared to the MBZ, which in our case is well-justified for twist angles larger than 0.5$^\circ$. In this case, we can discard umklapp processes.

\begin{figure}
\centering
\includegraphics[width=0.48 \textwidth]{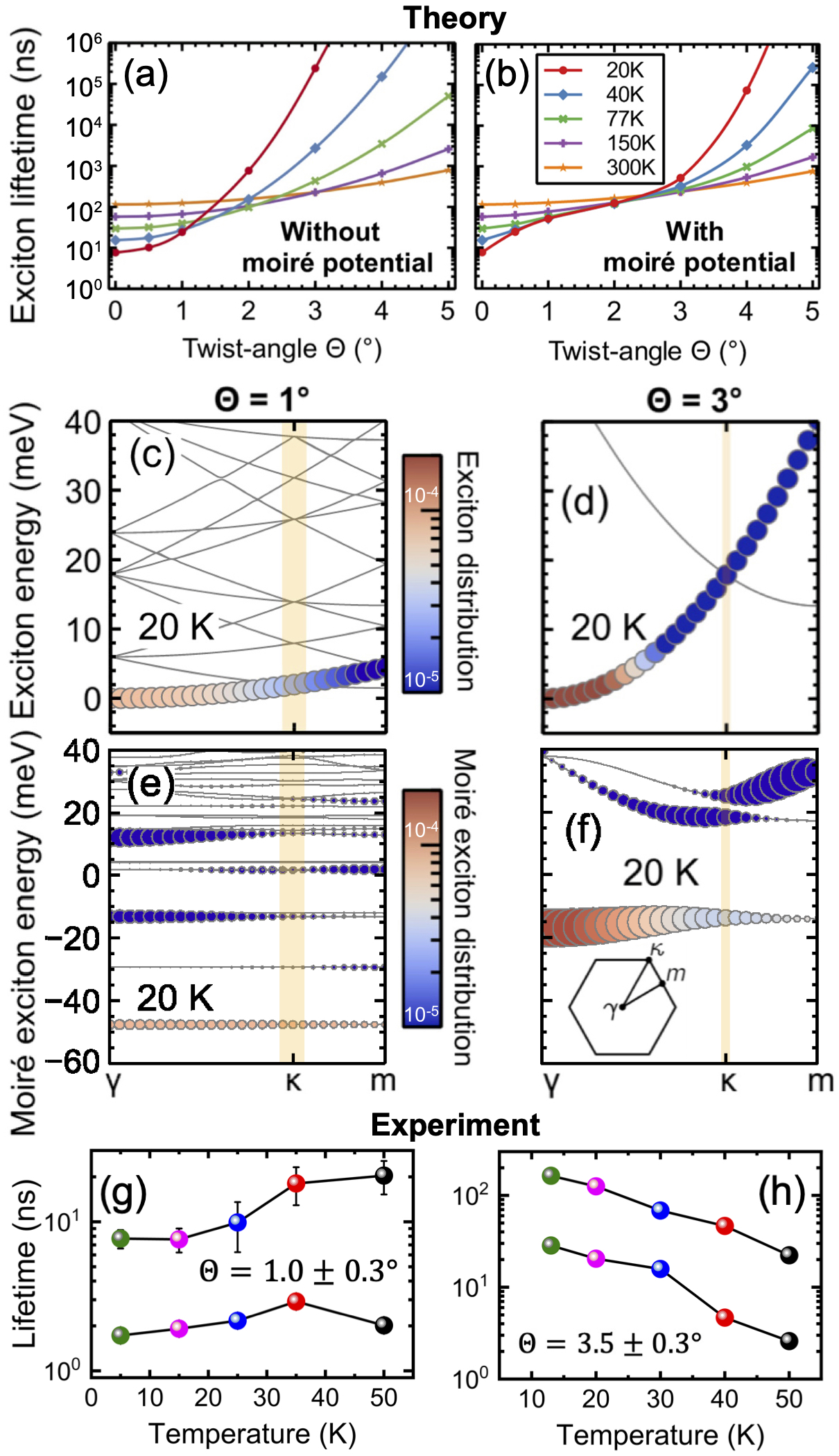}
\caption{\label{fig4}Temperature and twist angle dependence of calculated IX radiative lifetimes in the absence/presence of a moir\'{e} potential and measured temperature dependent IX dynamics. Calculated twist angle dependence of the IX lifetimes shown for various temperatures in the (a) \textit{absence} and (b) \textit{presence} of a moir\'{e} potential. Panels (c) and (d) show the IX band structure in the \textit{absence} of a moir\'{e} potential at twist angles of $\Theta =1^\circ$ and $\Theta=3^\circ$, respectively. Moir\'{e} exciton states are classified by momenta from the first MBZ shown in inset of (f). Light cone (orange shaded area) occupies a larger fraction of the MBZ for a TBL with a smaller twist angle. Panels (e)-(f) show the moir\'{e} exciton band structure, thermal distribution, and density-of-states contribution for different twist angles at low-temperature. Scale of the symbol size is the same for panel (c,d) and (e,f). Note that at small twist angle the IX bands become flat due to the moir\'{e} potential. (g, h) Measured temperature dependent fast and slow decay components of lifetimes near 1350 meV in the TBLs with $\Theta = 1.0 \pm 0.3 ^\circ$ ($3.5 \pm 0.3 ^\circ$) from 5 to 50 K (13 to 50 K). Error bars are smaller than the symbols for some data points.}
\end{figure}

In the numerical calculation shown in Fig.~\ref{fig4}, we assume the exciton population is thermalized obeying a Boltzmann distribution. To clarify the twist angle dependence, we disentangle the influence of the valley shift in reciprocal space from the energy modulation of moir\'{e} potential. First of all, a drastic change in IX lifetimes with the twist angle is expected even in the absence of a moir\'{e} potential by just considering the shifted valleys illustrated in Fig.~\ref{fig1}c and d. Only excitons within the light cone can decay radiatively. For a TBL with a small twist angle, low-temperatures are favorable to thermally keep as many excitons as possible within a narrow region around the $\Gamma$ point in the excitation picture. On the other hand, in a TBL with a large twist angle, light-matter coupling benefits from long tails in the high-temperature exciton distribution to overcome the large momentum shift $Q_{IX}$. As a result, the temperature dependence of IX lifetimes exhibits opposite behavior in TBLs with small and large twist angles. We therefore predict a crossover between these two regimes around an intermediate twist angle of $\theta=2^\circ$ as shown in Fig.~\ref{fig4}a.

Figure~\ref{fig4}c and d provide detailed insight for the cases of small and large twist angles at low-temperature, showing that at small twist angle the light cone overlaps with more occupied exciton states. Due to the absence of a moir\'{e} potential, the parabolic dispersion is simply folded back to the first MBZ without hybridization of bands. In Fig.~\ref{fig4}b, the presence of a moir\'{e} potential does not change the general trends but leads to a softening of the twist angle dependence. This can be understood as the moir\'e potential introducing an interaction between the IXs with different center-of-mass momenta as represented by the matrix form of eq.~\eqref{eq:moire_hamiltonian}. Thereby, the momentum conservation between the IX and photon in the light-matter coupling process is relaxed, opening additional channels for radiative recombination, as shown in Fig.~\ref{fig4}e-f. Correspondingly, the IX bands become less dispersive due to the moir\'e potential. Nevertheless, the IX lifetimes are predicted to change by several orders of magnitude when the twist angle is tuned between 1$^\circ$ and 3.5$^\circ$. Here, the symbol size quantifies the contribution of the moir\'{e} exciton states to the bright density-of-states as given by the coefficients $c^{\lambda}_{\bQ}$ in eq.~(\ref{eq:plane_wave}). Comparing panels e and f, we see that a small twist angle is particularly beneficial for light emission at low-temperatures due to the flattening of IX bands that leads to strong population of $\kappa$ excitons. We further compare the moir\'e exciton distribution at 20 K (Fig.~\ref{fig4}f) and 300 K (SM Fig.~S5b), demonstrating how a thermally broadened IX distribution fosters exciton-photon coupling at a large twist angle. In all cases, we find that higher excitonic bands become occupied  due to the moir\'{e}-induced mixture of bands represented by the relevant ($G_{\textrm{M}}=0$)-contributions ($c^{\lambda}_{\bQ}$ in eq.~(\ref{eq:plane_wave}) as shown in Fig.~\ref{fig4}e-f), in contrast to the cases when the moir\'e potential is not taken into account in Fig.~\ref{fig4}c-d.

To examine the predicted temperature dependence of IX radiative lifetimes, we compare measurements performed on two TBLs with the caveat that the total lifetime is measured. Figure ~\ref{fig4}g and h show the temperature dependent two decay times extracted by fitting with a bi-exponential function for the two TBLs with $\Theta$ = 1$^\circ$ and 3.5$^\circ$, respectively. Indeed, we experimentally observed opposite trends in the temperature dependent IX lifetimes in two TBLs. The IX lifetime increased (decreased) at higher temperature in the small (large) twist angle sample. This trend is consistent with the calculation (Fig. ~\ref{fig4}b) although a much weaker temperature dependence was observed. The deviation of the fast lifetime component from the $\Theta$ = 1$^\circ$ sample at $\sim$ 50 K may originate from increased phonon-mediated non-radiative processes.

We discuss the scope and limitations of our studies. First,  we focus on $\text{MoSe}_{\text{2}}$/$\text{WSe}_{\text{2}}$ TBLs with relatively small twist angles with an more accurate angle control than most previous experiments that have investigated the change of exciton resonances over a broader range of twist angle \cite{Zande:2014, Nayak:2017, Kunstmann:2018, Hsu:2019}. In TBLs with a small twist angle, the thermal distribution of the IX may be sufficient to satisfying the momentum conservation requirements of an indirect optical transition. In practice, phonon-assisted transitions do exist. Such transitions play an increasingly important roles in TBLs with larger twist angle ($\Theta> 5^\circ$). Second, the biexponential decay dynamics are not fully explained. We provide a rate equation analysis for a three-level system involving a dark state in the SM. Such a dark state may originate from an indirect transition involving electrons or holes in a different valley. Third, our calculations only address the radiative decay rate and our experiments measure the total decay rate. The calculation models an ideal situation where the IX dynamics is determined by the radiative process with unity quantum yield. It addresses the question of how radiative dacay can be controlled by the twist angle with and without the presence of an moir\'e superlattice. The qualitative agreement between the experiments and calculations is noteworthy. This agreement may suggest that  the non-radiative decay processes either do not depend on the twist angle strongly or depend on the twist angle in a similar way as the radiative decay. 

In conclusion, our studies represent a first step toward understanding how the twist angle of a vdW heterostructure can be used to control exciton dynamics. These results provide critical guidance to efforts on searching for exciton condensates where a long IX lifetime is preferred to reach an equilibrium temperature with the lattice. On the other hand, light-emitting devices should be built with TMD TBLs with a small twist angle. A faster radiative lifetime can compete favorably with other non-radiative processes and lead to higher quantum efficiency. Our work introduces new methods, “momentum-space band shift" and “periodic in-plane energy modulation" for controlling exciton lifetimes. These methods may be adapted by researchers investigating different classes of materials and be exploited by a wide variety of optoelectronic devices. \\

We gratefully acknowledge the helpful discussions with Kyounghwan Kim, Yimeng Wang, Wooyoung Yoon, and Emanuel Tutuc for sample preparation. The spectroscopic experiments performed by J.C. at UT-Austin and X. Li were primarily supported by NSF DMR-1808042. The support for K.T. was provided by the NSF MRSEC program DMR-1720595. Authors at UT-Austin acknowledge the use of facilities and instrumentation supported by the National Science Foundation through the Center for Dynamics and Control of Materials: an NSF MRSEC under Cooperative Agreement No. DMR-1720595. Material preparation was funded by the Welch Foundation via grant F-1662. R.C. was supported by NSF EFMA-1542747. K.W. and T.T. acknowledge support from the Elemental Strategy Initiative conducted by the MEXT, Japan ,Grant Number JPMXP0112101001,  JSPS KAKENHI Grant Numbers JP20H00354 and the CREST(JPMJCR15F3), JST. K. U. acknowledges support from the JSPS KAKENHI Grant No. 25107004. S.M. was funded for developing new CdSe architectures by the U.S. Department of Energy division of Energy Efficiency and Renewable Energy (EERE), grant no. M615002955. Work was performed in part at the Center for Integrated Nanotechnologies, a DOE, OBES Nanoscale Science Research Center \& User Facility, with aspects of the work supported by a CINT User Project 2017BRA0033. G.M. acknowledges support by the NSF Quantum Foundry through Q-AMASE-i program Award No. DMR-1906325. The theoretical calculations performed by M.F., A.S., D.E. and F.J were supported by the Deutsche Forschungsgemeinschaft (RTG 2247 ”Quantum Mechanical Materials Modelling”) 

\bibliography{references}
\end{document}

%\linespread{1.5}
\title{Supplementral Materials:\\
Twist Angle Dependent Interlayer Exciton Lifetimes in van der Waals Heterostructures}

\author{Junho Choi}
\thanks{J.C. and M.F. contributed equally to this work.}
\affiliation{Department of Physics and Center for Complex Quantum Systems, The University of Texas at Austin, Austin, TX 78712, USA.}
\author{Matthias Florian}
\thanks{J.C. and M.F. contributed equally to this work.}
\affiliation{Institute for Theoretical Physics, University of Bremen, 28334 Bremen, Germany.}
\author{Alexander Steinhoff}
\affiliation{Institute for Theoretical Physics, University of Bremen, 28334 Bremen, Germany.}
\author{Daniel Erben}
\affiliation{Institute for Theoretical Physics, University of Bremen, 28334 Bremen, Germany.}
\author{Kha Tran}
\affiliation{Department of Physics and Center for Complex Quantum Systems, The University of Texas at Austin, Austin, TX 78712, USA.}
\author{Dong Seob Kim}
\affiliation{Department of Physics and Center for Complex Quantum Systems, The University of Texas at Austin, Austin, TX 78712, USA.}
\author{Liuyang Sun}
\affiliation{Department of Physics and Center for Complex Quantum Systems, The University of Texas at Austin, Austin, TX 78712, USA.}
\author{Jiamin Quan}
\affiliation{Department of Physics and Center for Complex Quantum Systems, The University of Texas at Austin, Austin, TX 78712, USA.}
\author{Robert Claassen}
\affiliation{Department of Physics and Center for Complex Quantum Systems, The University of Texas at Austin, Austin, TX 78712, USA.}
\author{Somak Majumder}
\affiliation{Materials Physics \& Applications Division: Center for Integrated Nanotechnologies, Los Alamos National Laboratory, Los Alamos, New Mexico 87545, USA.}
\author{Jennifer A. Hollingsworth}
\affiliation{Materials Physics \& Applications Division: Center for Integrated Nanotechnologies, Los Alamos National Laboratory, Los Alamos, New Mexico 87545, USA.}
\author{Takashi Taniguchi}
\affiliation{International Center for Materials Nanoarchitectonics, National Institute for Materials Science,\\ 1-1 Namiki, Tsukuba, Ibaraki
305-0044, Japan.}
\author{Kenji Watanabe}
\affiliation{Research Center for Functional Materials, National Institute for Materials Science,\\ 1-1 Namiki, Tsukuba, Ibaraki
305-0044, Japan.}
\author{Keiji Ueno}
\affiliation{Department of Chemistry, Graduate School of Science and Engineering, Saitama University, Saitama, 338-8570, Japan.}
\author{Akshay Singh}
\affiliation{Department of Physics, Indian Institute of Science, Bengaluru, Karnataka 560012, India.}
\author{Galan Moody}
\affiliation{Department of Electrical and Computer Engineering, University of California Santa Barbara, Santa Barbara, CA 93106, USA.}
\author{Frank Jahnke}
\affiliation{Institute for Theoretical Physics, University of Bremen, 28334 Bremen, Germany.}
\author{Xiaoqin Li}
\thanks{Corresponding author}
\email{elaineli@physics.utexas.edu}
\affiliation{Department of Physics and Center for Complex Quantum Systems, The University of Texas at Austin, Austin, TX 78712, USA.}

%\setstretch{1.66}
\maketitle

\tableofcontents

%\clearpage
\section{S1. Details on sample preparation}
$\text{MoSe}_{\text{2}}$, $\text{WSe}_{\text{2}}$, and hexagonal boron nitride (hBN) layers were mechanically exfoliated from bulk crystals onto a polydimethylsiloxane (PDMS) sheet. The crystalline orientations of the $\text{MoSe}_{\text{2}}$ and $\text{WSe}_{\text{2}}$ monolayers were confirmed by second-harmonic generation (SHG) and rotationally aligned by high-resolution optical microscope images during the transfer. Each layer was transferred from the PDMS sheet to a targeted substrate or layer by 'stamping' dry-transfer method \cite{Andres:2014}. After each layer was transferred, thermal annealing in a high vacuum ($\sim10^{-6}$ mbar) at 200 $^{\circ}$C for 6 hours was performed to reduce the polymer residue from the surface contacted to the PDMS sheet.

\section{S2. SHG measurements on monolayers and a twisted bilayer (TBL)}
\begin{figure}[h!]
\centering
\includegraphics[width=0.35\textwidth]{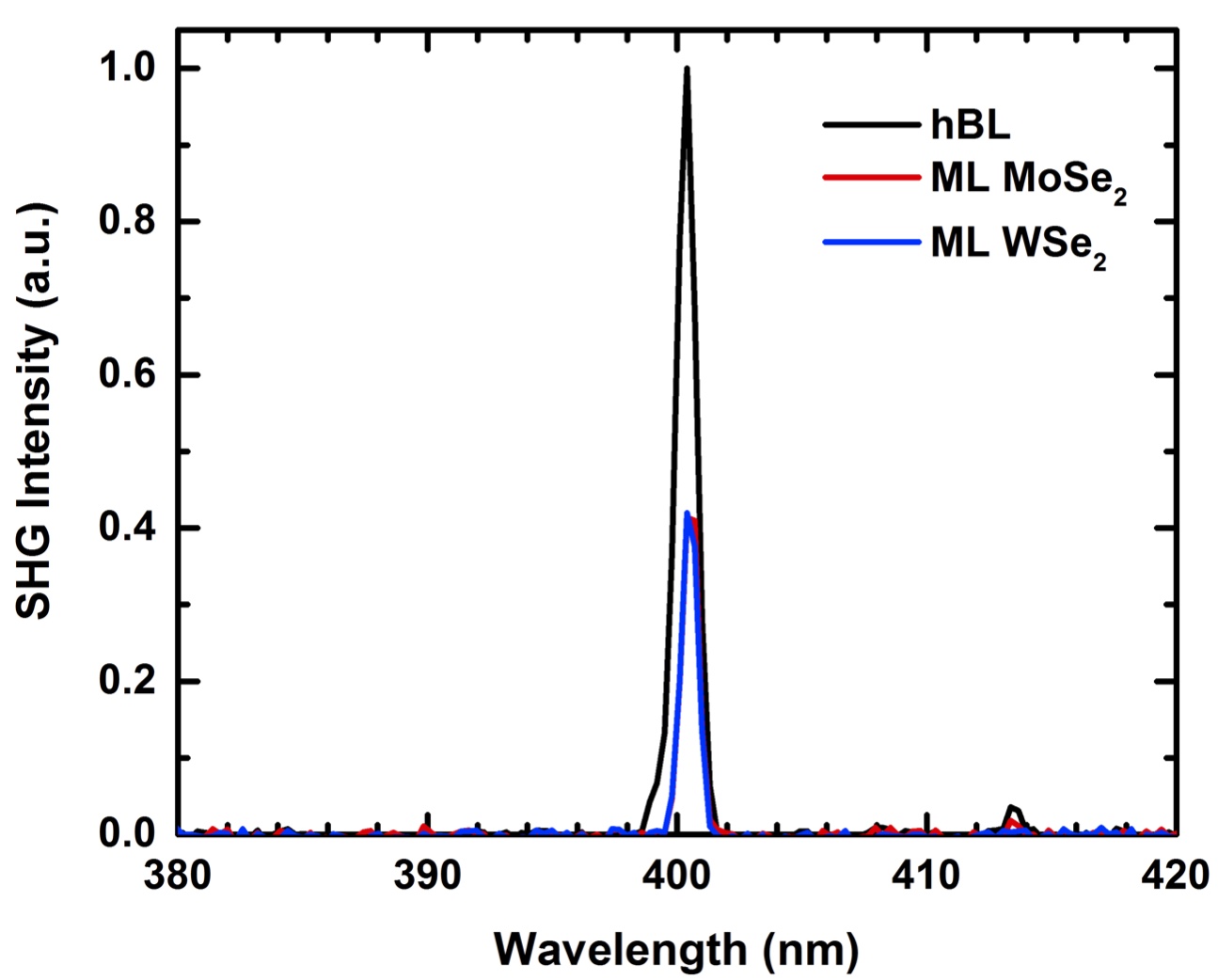}
\caption{SHG spectrum measured on $\text{MoSe}_{\text{2}}$ ML (red), $\text{WSe}_{\text{2}}$ ML (blue), and TBL region (black).}
\label{fig:SHG}
\end{figure}

The crystal orientations of $\text{MoSe}_{\text{2}}$ and $\text{WSe}_{\text{2}}$ monolayers (MLs) were determined by polarization-resolved SHG measurements. The details have been discussed in the previous studies \cite{Kumar:2013,Malard:2013,Li:2013}. Since the SHG measurements on both $\text{MoSe}_{\text{2}}$ and $\text{WSe}_{\text{2}}$ MLs exhibit a six-fold degeneracy, $R$- or $H$-stacking ($\Theta$ = 0$^\circ$ or 60$^\circ$) order of TBL region cannot be distinguished based on these measurements. In order to determine the relative twist-angle between the two MLs, SHG measurements on $\text{MoSe}_{\text{2}}$ ML, $\text{WSe}_{\text{2}}$ ML, and TBL regions were performed with a mode-locked ultrafast laser tuned to 800 nm. The SHG signals from individual MLs are expected to interfere with a relative phase difference depending on the twist-angle \cite{Hsu:2014}. The SHG intensity from TBL region with $R$- ($H$-) stacking order is expected be stronger (weaker) than ML regions. As shown in Fig.~\ref{fig:SHG}, the SHG signals on individual MLs show similar intensities while the SHG signal from the TBL region is enhanced compared to those from the ML regions, indicating the relative twist-angle between the MLs is near 0$^\circ$ stacking (or $R$-stacking).

\section{S3. Low-temperature photoluminescence (PL) measurements in TBLs}

For steady state PL measurements, a 660 nm continuous wave laser was focused to a spot size of $\sim$1 $\mu$m in diameter on the sample. The collected PL signal was guided to a spectrometer and detected by a charge-coupled device. All samples were held at $\sim$ 13 K and an average power of 100 $\mu$W was used to acquire the PL spectrum. For time-resolved PL (TRPL) measurements, 100 fs pulses at 680 nm with a 76 Mhz repetition rate generated by a Ti:sapphire-pumped optical parametric oscillator was used as the excitation source for a $\Theta = 1.0 \pm 0.3 ^\circ$ TBL. For TBLs with $\Theta = 2.2 \pm 0.3 ^\circ$ and $3.5 \pm 0.3 ^\circ$, 100 fs pulses tuned to 705 nm with a 76 Mhz repetition rate first pass through a pulse picker to reduce the repetition rate  in order to measure the longer lifetimes. The TRPL signal was sent through a spectrometer to select the energy of a particular interlayer exciton resonance. All TRPL measurements were performed at $\sim$ 13 K unless stated otherwise.

\begin{figure}[h]
\centering
\includegraphics[width=0.48\textwidth]{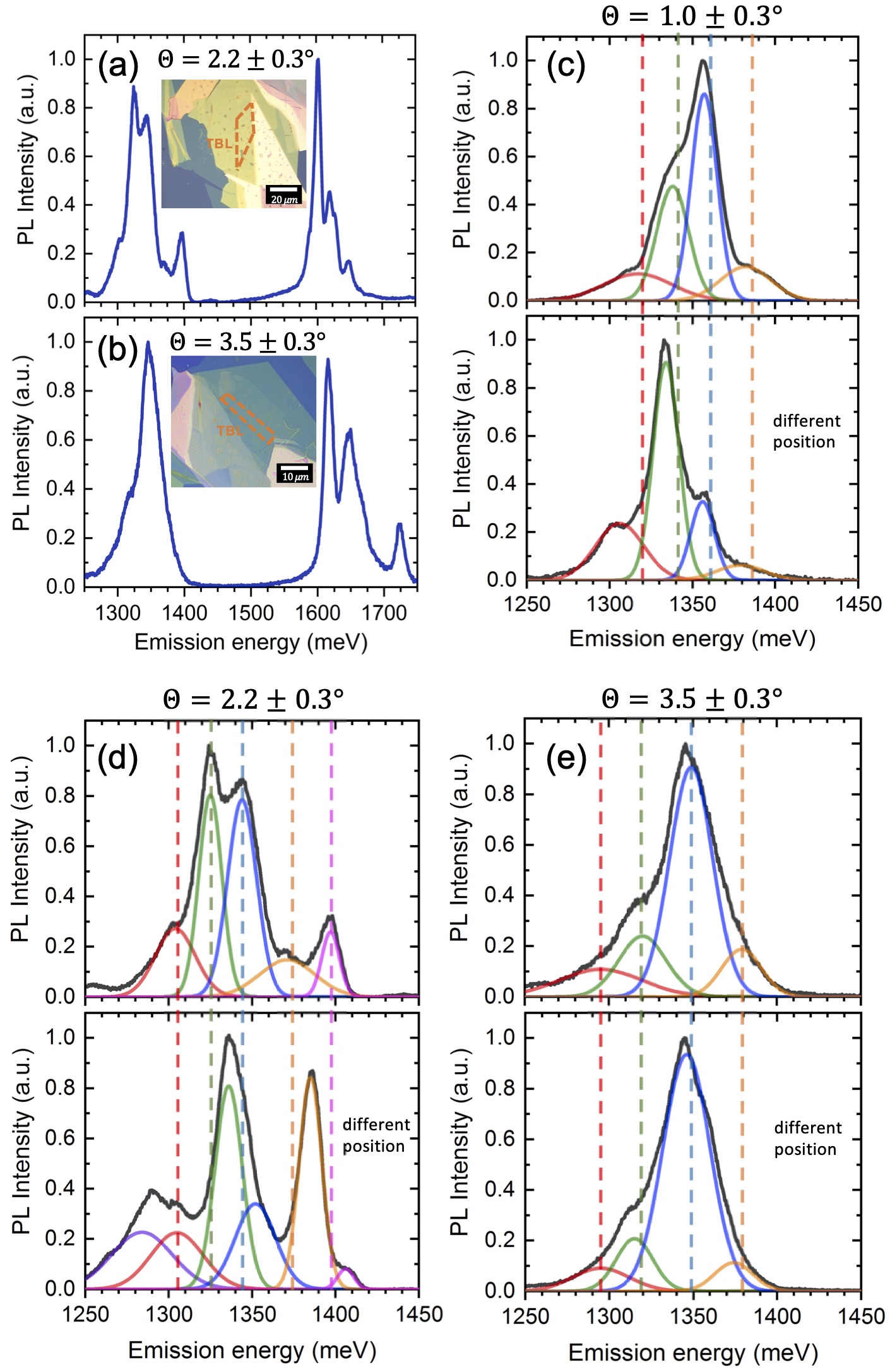}
\caption{Low-temperature PL spectrum of $\text{MoSe}_{\text{2}}$/$\text{WSe}_{\text{2}}$ TBLs and IX PL spectrum from different positions. (a, b) Low-temperature PL spectrum from hBN-encapsulated $\text{MoSe}_{\text{2}}$/$\text{WSe}_{\text{2}}$ TBLs with $\Theta = 2.2 \pm 0.3 ^\circ$ and $3.5 \pm 0.3 ^\circ$ twist angles over a broad spectral range to include both intra- and inter-layer excitons. Inset shows the optical image of the TBL with a scale bar and the TBL region is indicated with the orange dashed line. (c-e) Gaussian fitting to multiple IX resonances in the TBLs with three different twist angles and from two different positions on each sample TBL. Black solid line is the measured PL spectrum.}
\label{fig:PL}
\end{figure}

The TBL with $\Theta = 1.0 \pm 0.3 ^\circ$ twist-angle is prepared on sapphire substrate after encapsulation by hBN. The TBLs with $2.2 \pm 0.3 ^\circ$ and $3.5 \pm 0.3 ^\circ$ twist-angle are prepared on $\text{SiO}_{\text{2}}$/Si substrate, as shown in Fig.~\ref{fig:PL}. Different substrates can influence exciton resonant energy and radiative lifetime, the common bottom hBN layers reduce this dependence \cite{Raja:2019}. The thickness of all hBN layers for encapsulation is from 20 to 50 nm, providing the similar dielectric environment to all TBLs \cite{Gerber:2018}. In low-temperature PL spectrum, both intra- and inter-layer exciton (IX) resonances are clearly observed in all TBLs. The microscope images and PL spectrum for the $\Theta = 1.0 \pm 0.3 ^\circ$ are shown in the main text while those for the other two TBLs are shown in Fig.~\ref{fig:PL}. The averaged energy spacings are 21 $\pm$ 2, 23 $\pm$ 2, and 28 $\pm$ 3 meV for $\Theta$ = 1, 2.2, and 3.5$ ^\circ$ TBLs, respectively. The observation of larger energy spacing with increasing the twist angle is consistent with a smaller lateral size of the quantum confinement imposed by the moir\'{e} supercell. We compare the IX PL spectra from two different positions on the same TBL as shown in Fig.~\ref{fig:PL}c-e, for all three TBLs. We consistently observe multiple IX resonances and the energies of IXs shift $\sim$10 meV. The slight energy variation of IX resonances may be resulting from the dielectric disorder due to the residue. A similar amount of spectral shift has been reported in other experiments \cite{Raja:2019}.

\section{S4. Twist angle dependent moir\'e superlattice periodicity}
\begin{figure}[h!]
\centering
\includegraphics[width=0.28\textwidth]{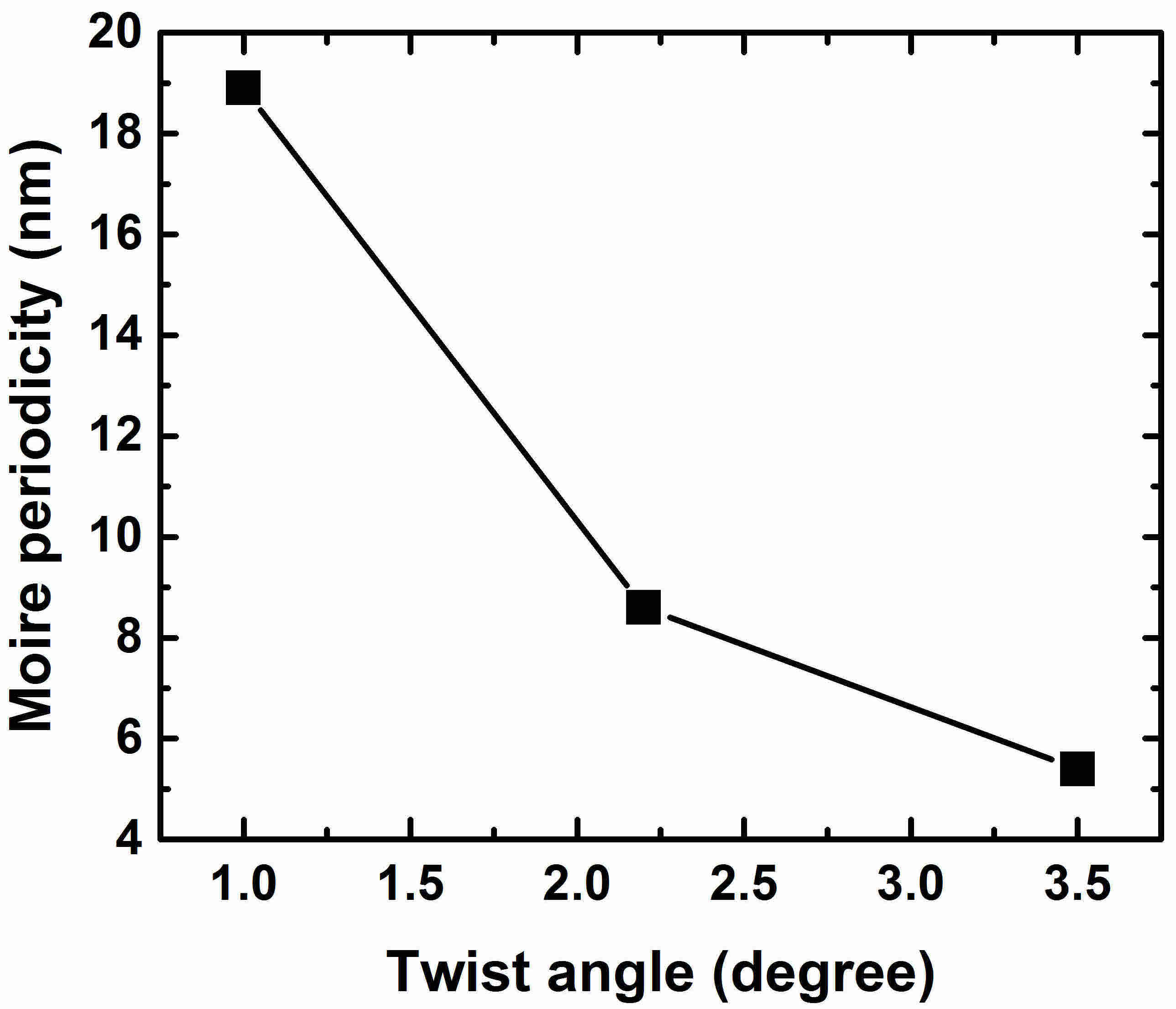}
\caption{Twist angle dependent moir\'e periodicity of the TBLs with $\Theta$ = 1$^\circ$ to 3.5$^\circ$.}
\label{fig:periodicity}
\end{figure}

The lattice constants of $\text{MoSe}_{\text{2}}$ and $\text{WSe}_{\text{2}}$  ($\simeq  0.1 \% $) are very similar because of the common chalcogen (Se) atom. The periodicity ($a_M$) of moir\'e pattern in $\text{MoSe}_{\text{2}}$/$\text{WSe}_{\text{2}}$ TBL with a small relative twist angle ($\Theta$) can be estimated by \cite{Wu:2018},
\begin{equation} \label{moire_eqn}
a_M \approx a_0/\sqrt{\Theta^2+\delta^2}
\end{equation}
where the lattice mismatch ($\delta=|a_0-a_0'|/a_0$) is given by the lattice constants of two MLs ($a_0$ and $a_0'$). For TBLs with $\Theta$ = 1$^\circ$ - 3.5$^\circ$, the contribution of $\delta$ is negligible. In other words, the moir\'e superlattice periodicity is primarily controlled by the twist angle between the two MLs. The moir\'e periodicity is changed from 18.9 to 5.4 nm when the twist-angle is changed from 1$^\circ$ to 3.5$^\circ$, as shown in Fig.~\ref{fig:periodicity}.

\section{S5. Estimation of IX quantum yield at low-temperature under non-resonant excitation}
Under resonant excitation conditions, the quantum yield ($\eta$) of a two-level emitter can be defined in terms of radiative decay ($\gamma_{rad}$) and non-radiative decay ($\gamma_{nrad}$) rates according to the relation, $\eta=\gamma_{rad}/(\gamma_{rad}+\gamma_{nrad}) = \tau_{total}/\tau_{rad}$ where the total decay rate is defined as $\gamma_{total} = \gamma_{rad}+\gamma_{nrad} = 1/\tau_{total}$.

It is extremely challenging to obtain such measurements for the IXs because of the small dipole moments and weak absorption. We measured the relative PL quantum yield (QY) of IXs under non-resonant excitation conditions, which serves to provide an empirical guidance of the expected PL signal. We used a thin film of diluted colloidal CdSe/CdS core/shell quantum dots (QDs), comprising an ultra-thick CdS shell ($\sim$16 monolayers), with near 50$\% $ QY as a calibration sample \cite{Chen:2008, Orfield:2018}. We used spin coating to prepare multiple films with different concentrations of diluted QD solution and measured the absorption and PL under the same experimental conditions. The area of PL spectrum was integrated to count the total number of emitted photons. We confirm that there is linear relation between the PL counts as a function of absorbance, indicating no re-absorption effect of emitted photons in the calibration samples, as expected for the thick-shell QDs for which the predominant absorption onset occurs above 2.4 eV; whereas, emission is at $\sim$1.9 eV \cite{Hanson:2015}. The TBL with $\Theta = 2.2 \pm 0.3 ^\circ$ twist-angle was chosen as a representative sample and the QY at low-temperature was estimated to be $\simeq  0.1 \% $. This relatively low QY under the non-resonant excitation conditions can be attributed to different decay channels including the population of dark states, an efficient radiative decay of the intralayer excitons, and phonon assisted non-radiative decays.

\section{S6. Measured energy resolved TRPL}
\begin{figure}[h]
\centering
\includegraphics[width=0.48\textwidth]{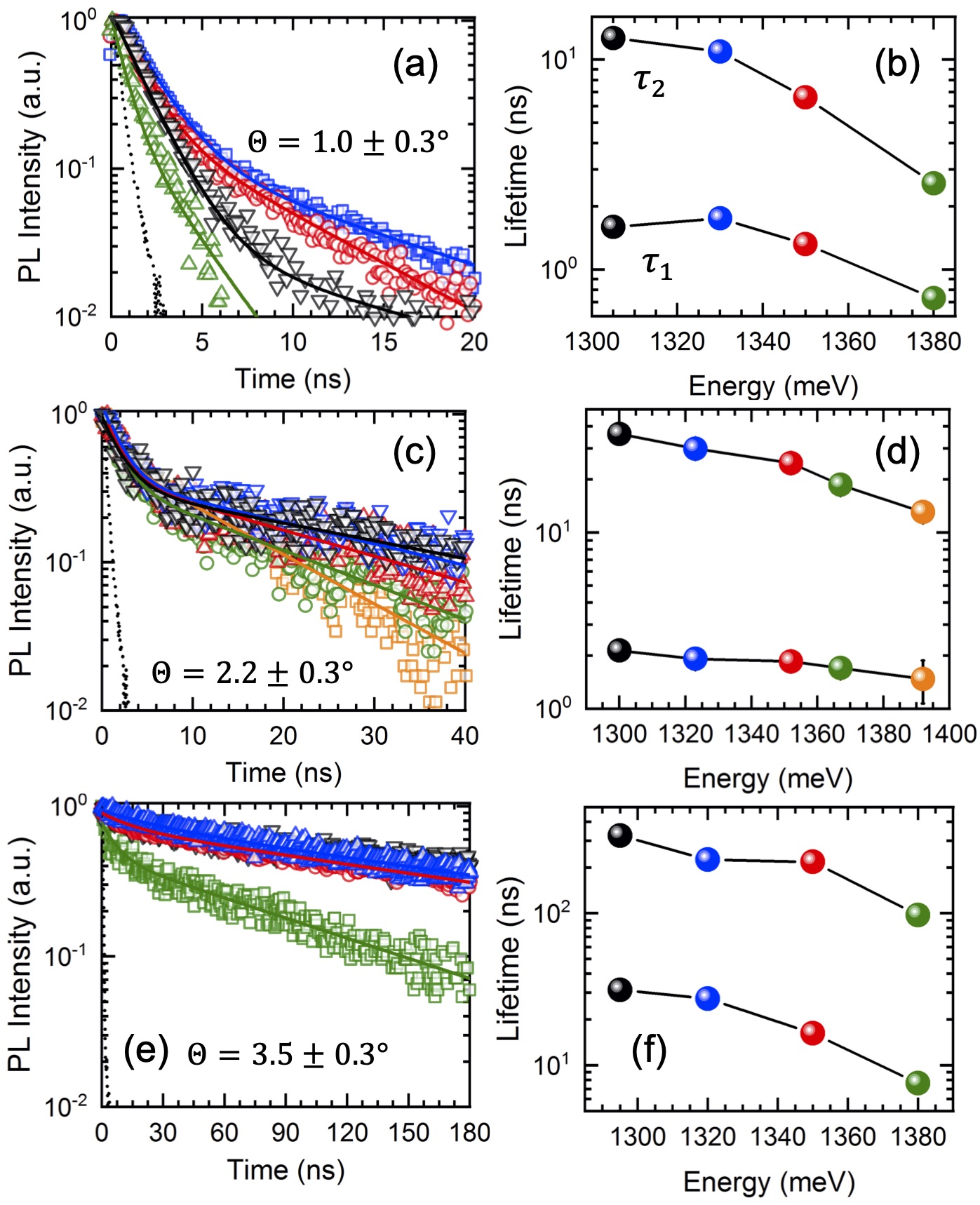}
\caption{Emission energy resolved TRPL in TBLs with (a) $\Theta = 1.0 \pm 0.3 ^\circ$, (c) $2.2 \pm 0.3 ^\circ$, and (e) $3.5 \pm 0.3 ^\circ$. (b, d, f) Fast ($\tau_{1}$) and slow ($\tau_{2}$) component of IX lifetime extracted by fitting with biexponential function for TBL with these three different twist angles.}
\label{fig:E_lifetime}
\end{figure}

Emission energy resolved TRPL measured from three different twist angles are presented in Fig.~\ref{fig:E_lifetime}. We include all interlayer exciton resonances here. For all IX resonances, we observe the consistent trend that a longer IX lifetime is found in a TBL with a larger twist angle than that in a TBL with a smaller twist angle .

\section{S7. Calculated moir\'e exciton band structure at 300 K and measured temperature dependent TRPL spectrum}
\begin{figure}[h]
\centering
\includegraphics[width=0.48\textwidth]{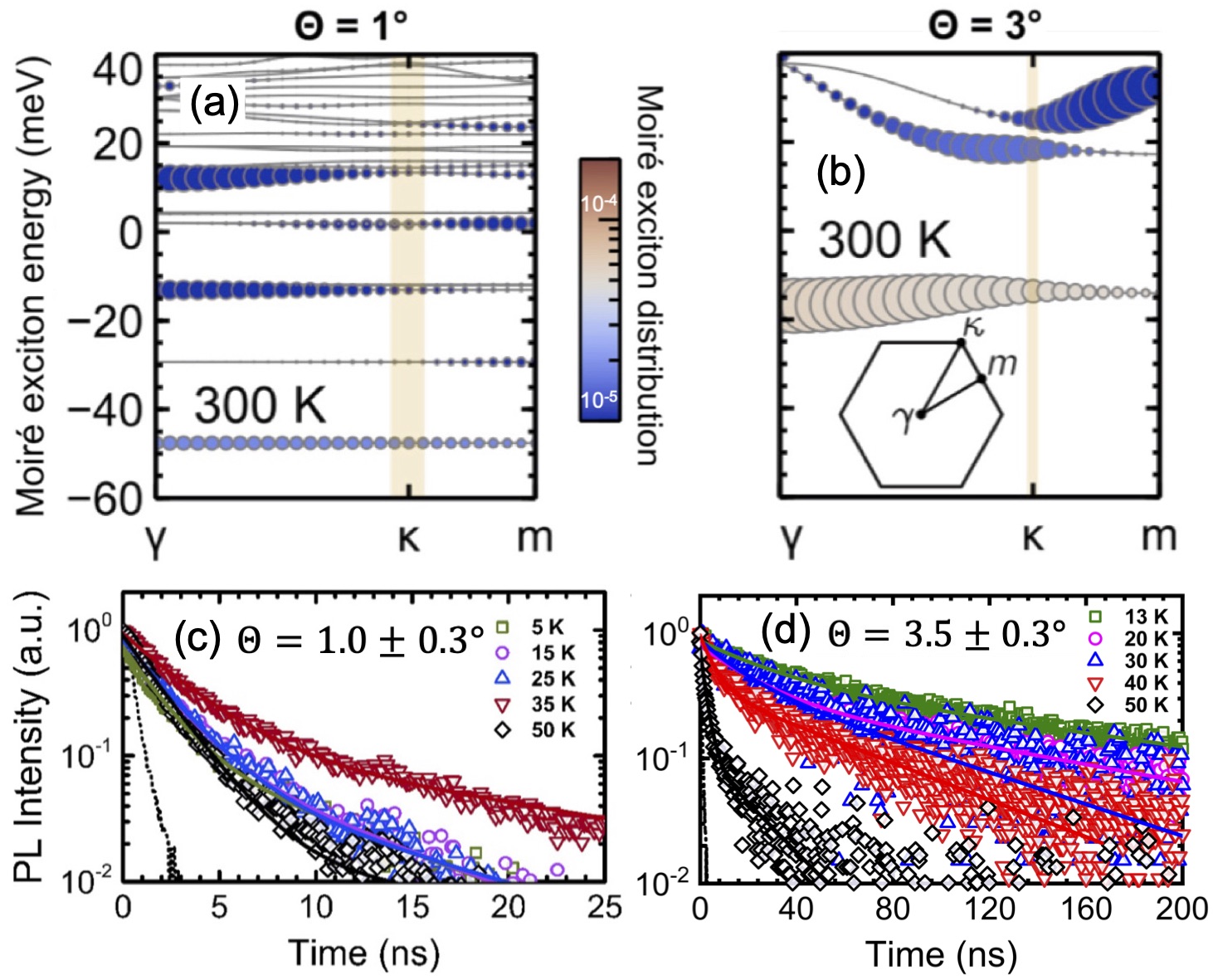}
\caption{Moir\'e exciton band structure at 300 K in TBLs with twist angles of (a) $\Theta =1^\circ$ and (b) $\Theta=3^\circ$, respectively. Inset in panel (b) illustrates the first moir\'{e} Brillouin zone (MBZ). The scale of the symbol size and color scale are the same as Fig. 4 in the main text. Panels (c) and (d) show temperature dependent TRPL spectrum from TBLs with $\Theta = 1.0 \pm 0.3 ^\circ$ and $3.5 \pm 0.3 ^\circ$.}
\label{fig:TD_moire}
\end{figure}

Figure~\ref{fig:TD_moire} shows the IX band structure in the presence of moir\'e potential at 300 K and temperature dependent TRPL spectrum from TBLs with two different twist angles. We observe that the thermally broadened IX distribution leads to a significantly stronger exciton-photon coupling at large twist angles compared to the low-temperature results shown in Fig. 4 in the main text.

\section{S8. Rate equations for a three-level system including a dark exciton state}
\begin{figure}[h]
\centering
\includegraphics[width=0.28\textwidth]{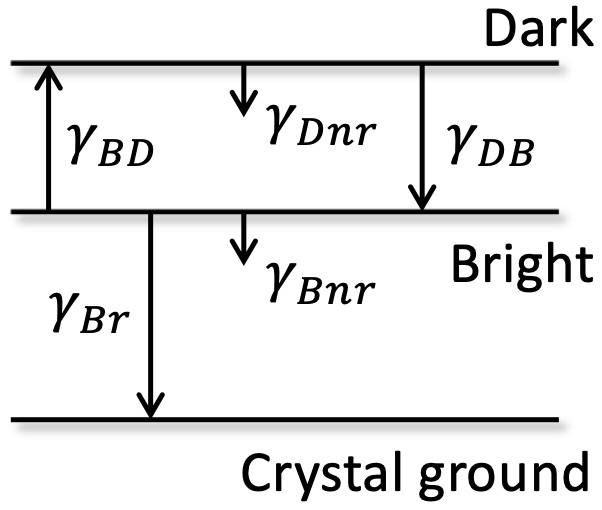}
\caption{Schematic of three level systems: dark exciton, bright exciton, and crystal ground state. The bright exciton can radiatively decay ($\gamma_{Br}$) and both bright and dark excitons can non-radiatively decay ($\gamma_{Bnr}$ and $\gamma_{Dnr}$). The bright and dark exciton states can be converted ($\gamma_{BD}$ and $\gamma_{DB}$) to the other state via thermal or valley scattering processes.}
\label{fig:rate_equation}
\end{figure}

To describe the biexponential decay observed, we model the decay dynamics of a three-level system consisting of the bright exciton, dark exciton, and crystal ground state, often used in modeling semiconductor nanostructures \cite{Dalgarno:2005}. The relevant dynamic parameters include the bright excition radiative decay rate ($\gamma_{Br}$), non-radiative decay rates ($\gamma_{Bnr}$ and $\gamma_{Dnr}$), and conversion rates between the bright and dark exciton states ($\gamma_{BD}$ and $\gamma_{DB}$), as shown in Fig.~\ref{fig:rate_equation}. The rate equations are 
\begin{eqnarray}
    \begin{split}
       &\frac{dn_{B}}{dt} = -n_{B}(\gamma_{Br}+\gamma_{Bnr}+\gamma_{BD})+n_{D}\gamma_{DB}\ ,\\
       &\frac{dn_{D}}{dt} =-n_{D}(\gamma_{Dnr}+\gamma_{DB})+n_{B}\gamma_{BD}\ ,
    \end{split}
\end{eqnarray}
where $n_{B}$ and $n_{D}$ are occupations in the bright and dark state, respectively. The solution of the rate equations is a biexponential decay function, $n_{B}(t)=a_{1}e^{-\gamma_{1}t}+a_{2}e^{-\gamma_{2}t}$, with the fast and slow decay components
\begin{eqnarray}
\begin{split}
\gamma_{1,2}&=\frac{1}{2}(\gamma_{Br}+\gamma_{Bnr}+\gamma_{Dnr}+\gamma_{BD}+\gamma_{DB})\\
&\pm\frac{1}{2}\sqrt{(\gamma_{Br}+\gamma_{Bnr}+\gamma_{Dnr}+\gamma_{BD}+\gamma_{DB})^2} \\ 
&\overline{-4\gamma_{r}(\gamma_{Dnr}+\gamma_{DB})-4(\gamma_{Bnr}\gamma_{Dnr}+\gamma_{BD}\gamma_{Dnr}+\gamma_{DB}\gamma_{Bnr}}\ .
\end{split}
\end{eqnarray}
In the case of $\gamma_{Br}\gg\gamma_{BD},\gamma_{DB}$, the fast and slow decay rates can be approximated as
\begin{eqnarray}
    \begin{split}
        &\gamma_{1}\simeq\gamma_{Br}+\gamma_{Bnr}+\gamma_{BD}\ ,\\
        &\gamma_{2}\simeq\gamma_{Dnr}+\gamma_{DB}\ .
    \end{split}
\end{eqnarray}
Using $\gamma_{1}$ ($= 1/\tau_{1}$) and $\gamma_{2}$ ($= 1/\tau_{2}$) to represent the fast and slow decay component, this model may be used to explain the biexponential decay observed in our TRPL experiments by attributing the fast (slow) decay component to the bright (dark) exciton lifetime.

\section{S9. Moir\'e exciton lifetime calculations}

The radiative moir\'e exciton recombination rate is calculated using the Fermi's golden rule~\cite{esser_photoluminescence_2000}
\begin{eqnarray}
\tau^{-1}_{\text{IX}} = \frac{2\pi}{\hbar} \sum\limits_{if}| \bra{f} H_{\text{rad}} \ket{i} |^2 \delta(\varepsilon_i - \varepsilon_f) N_i\,
\label{eq:IX_fermisgoldenrule0}
\end{eqnarray}
where the optical matrix element describes transitions between the initial exciton state $\ket{i}$ and the final photon state $\ket{f} = b^\dagger_{\bq, \sigma}\ket{0}$. Here, $b^\dagger_{\bq, \sigma}$ creates a photon with wave vector $\mathbf q$ and polarization $\sigma$ with respect to the vacuum state $\ket{0}$. $N_i$ is the occupation of the initial exciton states.\\ Under weak excitation conditions and using the Coulomb-gauge the light-matter interaction can be written as
\begin{eqnarray}
    \begin{split}
        H_{\text{rad}} &=& \frac{1}{c}\sum_{\bR\bR'\nu\nu'}\mathbf{A}\left(\frac{\bR+\bR'}{2}\right) {a}^{\dagger,\nu}_{\bR}{a}^{\nu'}_{\bR'}\times \\
        &&\times\int\,\text{d}^3r\,u^*_{\nu}(\mathbf{r}-\bR)\,\mathbf{j}\,u_{\nu'}(\mathbf{r}-\bR')\,,
        \label{eq:Hrad}
    \end{split}
\end{eqnarray}
where ${a}^{\dagger,\nu}_{\bR}$ creates an electron at lattice site $\bR$ in band $\nu$. To obtain eq.~\eqref{eq:Hrad}, we have assumed that the Wannier functions $u_{\nu\sigma}(\mathbf{r}-\bR)$ are localized at the lattice sites and that the vector potential $\mathbf{A}(\bR)$ is slowly varying over the unit cell.

The vector potential is expanded into a plane-wave basis 
\begin{eqnarray}
\mathbf{A}(\bR) = \sum_{\bq\sigma} \sqrt\frac{2\pi\hbar c}{n^2 q \Omega}\mathbf{e}_{\bq\sigma}\left( b_{\bq, \sigma} e^{i \bq_\parallel \bR}  + b^\dagger_{-\bq, \sigma} e^{-i \bq_\parallel \bR}\right)\,,
\end{eqnarray}
where $\Omega$ is the normalization volume of the photon eigenmodes, $\mathbf{e}_{\bq\sigma}$ are the unit vectors of the two transversal photon polarizations and $n$ is the average refractive index of the dielectric environment of the heterobilayer.
For the optical matrix element in eq.~\eqref{eq:Hrad} we use the two-center approximation~\cite{bistritzer_moire_2011}. As a result the matrix element
\begin{eqnarray}
    &&\int\,\text{d}^3r\,u^*_{\nu}(\mathbf{r}-\bR)\,\mathbf{j}\,u_{\nu'}(\mathbf{r}-\bR') \approx \mathbf{j}_{\nu\nu'}(\bR-\bR')
    \label{eq:two_center}
\end{eqnarray}
depends only on the difference between the electron and hole position and can be expressed in terms of its Fourier transform 
\begin{eqnarray}
\mathbf{j}_{\nu\nu'}(\bR-\bR') = \frac{1}{N}\sum_{\bq}e^{-i\bq(\bR-\bR')}\mathbf{j}_{\nu\nu'}(\bq)\,,
\end{eqnarray}
where $N$ is the number of unit cells in each layer.

To describe the moir\'{e} exciton states we use a low energy continuum model that we derive along the lines of Ref.~\cite{Yu:2015, Wu:2017, Yu:2017, Wu:2018}. For each monolayer the band edge electrons and holes at the K point are predominantly described by Wannier functions of the metal d-orbitals~\cite{liu_three-band_2013}. In the absence of interlayer hybridization, the IX is described by the wave function
\begin{eqnarray}
\ket{\bQ} &=& \frac{1}{\sqrt{\mathcal{A}}}\sum\limits_{\bk} \phi_X(\bk) {a}^{\dagger,c}_{\bK_c+\bk + \frac{m_e}{M}\bQ} {a}^v_{\bK_v+\bk-\frac{m_h}{M}\bQ} \ket{0} \label{eq:X_wfct}
\end{eqnarray}
with eigenenergies $E(\bQ)= \hbar^2|\bQ|^2/(2M) + E_{\text{gap}} - E_{B}$ that are characterized by the center-of-mass momentum $\bQ$, $M=m_e+m_h$ is the total exciton mass, and $m_e$ ($m_h$) is the electron (hole) effective mass. $\mathcal{A}$ is the crystal area. The operator ${a}^{\dagger,c}_{\bK_c + \bq} ({a}^{v}_{\bK_v + \bq})$ creates (annihilates) an electron in the K$_c$ (K$_v$)-valley of the conduction (valence) band and can be expanded in terms of their Wannier counterparts
\begin{eqnarray}
    {a}^{\dagger,c}_{\bq} = \frac{1}{\sqrt{N}}\sum_{\bR} e^{ i\bq\bR} {a}^{c}_{\bR} \quad\text{and}\quad
    {a}^{v}_{\bq} = \frac{1}{\sqrt{N}}\sum_{\bR} e^{-i\bq\bR} {a}^{v}_{\bR}\,.
\end{eqnarray}
 The electron-hole relative-motion wave function in momentum space $\phi_X(\bk)$ is determined by the Wannier equation 
\begin{eqnarray}
\sum_{\bq'}\left[\frac{\hbar^2\bq^2}{2\mu}\delta_{\bq\bq'} - \frac{1}{\mathcal{A}}V_{|\bq-\bq'|}\right]\phi_X(\bq') = -E_{B}\phi_X(\bq)
\label{eq:wannier_eq}
\end{eqnarray}
where $\mu=m_em_h/(m_e+m_h)$ is the reduced mass. The solution of eq.\eqref{eq:wannier_eq} determines the exciton energy $E_{\text{gap}} - E_{B}$, where $E_{B}$ is the exciton binding energy and $E_{\text{gap}}$ the band gap. For the effective masses we use $m_e = 0.57 m_0$ and $m_h = 0.36 m_0$, where $m_0$ is the free electron mass.~\cite{wang_optical_2019} Details about the dielectrically screened Coulomb potential are given in (S7). We find that the IX binding energy $E_B$ for the hBN-encapsulated $\text{MoSe}_{\text{2}}$/$\text{WSe}_{\text{2}}$ heterobilayer is about $110$ meV. The exciton Bohr radius $a_B$ can be approximated by the root mean square of the electron hole separation and we obtain a value of $1.15$ nm.

An interlayer twist in real space generates a relative shift in momentum space ($\bK_c - \bK_v$) as well as a spatial modulation of the exciton energy. For small twist angle, the moir\'{e} periodicity can be assumed to be large compared to the exciton Bohr radius. Thus, we neglect the variation of the binding energy in the moir\'{e} pattern and use a local approximation of the exciton moir\'{e} potential $V^{\textrm{M}}(\bR)$ according to Ref.~\cite{Wu:2018}. In this framework, the exciton Hamiltonian reads
\begin{eqnarray}
H &=& -\frac{\hbar^2}{2 M}\Delta_{\bR} + V^{\textrm{M}}(\bR)
\label{eq:moire_hamiltonian}
\end{eqnarray}
and can be diagonalized using a plane-wave expansion 
\begin{eqnarray}
&&\left[\frac{\hbar^2}{2M}\left(\bQ - \bG_{\text{M}}\right)^2 - \varepsilon_{\bQ,\lambda}\right] c^{\lambda }_{\bQ-\bG_{\text{M}}}\delta_{\bG_{\text{M}},\bG_{\text{M}}'} + \nonumber\\
&&+ \sum_{\bG_{\text{M}}'} V^{\textrm{M}}_{\bG_{\text{M}},\bG_{\text{M}}'} c^{\lambda }_{\bQ-\bG_{\text{M}}'} = 0
\end{eqnarray}
with eigenenergies $\varepsilon_{\bQ,\lambda}$ and eigenstates
\begin{eqnarray}
\ket{ \bQ, \lambda } = \sum\limits_{\bG_{\text{M}}} c^{\lambda }_{\bQ-\bG_{\text{M}}} \ket{\bQ - \bG_{\text{M}}}\,.
\label{eq:moire_eigenstates}
\end{eqnarray}
The moir\'{e} reciprocal lattice vectors $\bG_{\text{M}}$ are derived as differences of top and bottom layer reciprocal lattice vectors. Therefore, the size of the MBZ scales with the interlayer twist angle. Following Ref.~\cite{Wu:2018} the moir\'e potential can be parameterized according to 
\begin{eqnarray}
    V^{\textrm{M}}(\bR)\approx\sum_{j=1}^6 V_{j} e^{i\bG_{\text{M},j}\bR}\,,
\end{eqnarray}
where the Fourier series is restricted to the first shell of reciprocal Moire lattice vectors $\bG_{\text{M},j}$. Due to $\hat C_3$ symmetry $V_1=V_3=V_5$, as well as $V_2=V_4=V_6$. Because $V^{\textrm{M}}(\bR)$ is real, $V_1=V_4^*\equiv V\exp(-i\psi)$. For the parameter $(V,\psi)$ we use $(11.8, 79.5^\circ)$ as obtained from DFT calculations for AA stacked MoSe$_2$/WSe$_2$ heterobilayer~\cite{Wu:2018}.

Using the above plane-wave expansion, the optical matrix element between the initial moir\'{e} exciton state $\ket{\bQ,\lambda}$ and the final photon state $\ket{\bq,\sigma}$ can be expressed as:
\begin{eqnarray}
&&\bra{\bq,\sigma}H_{\text{rad}}\ket{\bQ,\lambda} = \sqrt{\frac{2\pi\hbar}{c n^2 |\bq| \Omega}}\,  \phi_X(\br=0) \sum\limits_{\bG_{\text{M}},\bG_c,\bG_v} c^\lambda_{\bQ+\bG_{\text{M}}}\times\nonumber\\
&&\times\,\mathbf{j}_{\text{vc}}\left((\bK_{v} + \bG_{v}) -\frac{m_h}{M}(\bQ - \bG_{\text{M}}) + \frac{\bq_{\parallel}}{2} \right)\cdot \mathbf{e}_{\bq\sigma} \times \nonumber\\ 
&&\times\,\delta_{\bQ-\bG_{\text{M}}, ( \bK_c +\bG_c ) - ( \bK_v + \bG_v ) + \bq_\parallel} \,.
\label{eq:Hrad_IX_final}
\end{eqnarray}
Here, the interlayer current matrix element $\mathbf{j}_{\text{vc}}(\bQ)$ is expected to rapidly decline on the scale of the monolayer lattice vector~\cite{Wu:2018}. We assume that the twist angle is sufficiently large so that typical photon momenta are still small compared to the MBZ, which in our case is well-justified for twist angles larger than 0.5$^\circ$. In this case, umklapp processes can be discarded. As a result, moiré excitons can only recombine if the momentum conservation law $\bQ - \bq_\parallel= \bK_c - \bK_v := \kappa$ is fulfilled where $\bq_\parallel$ represents the in-plane component of the photon wave vector. Under these assumptions the optical matrix element reads
\begin{eqnarray}
&&\bra{\bq,\sigma}H_{\text{rad}}\ket{\bQ,\lambda}\approx
\sqrt{\frac{2\pi\hbar}{c n^2 |\bq| \Omega}}\,  \phi_X(\br=0) \,c^\lambda_{\bQ} \times\nonumber \\&&\times\,\mathbf{j}_{\text{vc}} \cdot \mathbf{e}_{\bq\sigma}\,\delta_{\bQ, \bK_c - \bK_v + \bq_\parallel} \delta_{\bG_{\text{M}},0} \, ,
\end{eqnarray}
where $\mathbf{j}_{\text{vc}}=\bra{v\bK}\mathbf{j}\ket{c\bK}$ is the interlayer current matrix element of the untwisted heterobilayer. Optical transitions at the $\mathbf{K}$ point satisfy circular polarization selection rules with matrix elements $|\mathbf{j}_{\text{vc}}\cdot \mathbf{e}^*_+| / \mathcal{D} = 0.22$ and $|\mathbf{j}_{\text{vc}}\cdot \mathbf{e}^*_-| / \mathcal{D} = 0.06$, where $\mathbf{e}_\pm=(1,\pm i)/\sqrt 2$ and $\hbar/e\mathcal{D} =$ 4.43 eV $\text{\normalfont\AA}$. ~\cite{Wu:2018}. We neglect the significantly weaker transition with $\mathbf{e}_-$ polarization direction.

We decompose the photon wave vector into the in- and out-of-plane components as $\bq=\{\bq_\parallel, q_z\}$ such that the summation over $q_z$ in Eq.~\eqref{eq:IX_fermisgoldenrule0} yields the one-dimensional density-of-states and gives rise to radiative decay of excitons with wave vectors inside the light cone $|\kappa| - q_\lambda < |\bQ| < |\kappa| + q_\lambda$. The boundary of the light cone is determined by the photon dispersion $\hbar c\, q_\lambda = n\,\varepsilon_{\kappa,\lambda}$.
Combining eqs.~\eqref{eq:IX_fermisgoldenrule0} and \eqref{eq:Hrad_IX_final} we finally obtain for the moiré exciton lifetime
\begin{eqnarray}
&&\tau^{-1}_{\text{IX}} = \frac{2\pi}{\hbar c^2} g^{\text{rad}}|\phi_X(\br=0)|^2 \sum\limits_{{\lambda}} \int \frac{d^2 {\bq}_\parallel}{(2 \pi)^2}   \Big| j_{\text{vc}} c^{\lambda}_{\kappa+{\bq}_\parallel} \Big|^2 N_{\kappa + {\bq}_\parallel}^{\lambda} \times \nonumber \\
&& \times \frac{1}{q_{\lambda}}\left\{ \frac{1}{\sqrt{1 - \left(q_\parallel/q_{\lambda}\right)^2} } - \sqrt{1 - \left(q_\parallel/q_{\lambda}\right)^2} \right\} \Theta{ ( q_{\lambda} - q_{\parallel}) } \, . \label{eq:lifetime}
\end{eqnarray}
Here, $g^{\text{rad}}$ is the degeneracy of optically active exciton states and accounts for time-reversal symmetry. The moiré exciton distribution $N_{\bQ}^{\lambda}$ is assumed to be thermalized obeying a Boltzmann distribution. We consider the low density limit such that the exciton distribution is normalized to unit total density, $g_X\sum_{\bQ,\lambda} N_{\bQ}^{\lambda} = 1$, where $g_X$ denotes the degeneracy of occupied (bright and dark) exciton states. In particular, in the limit of zero twist-angle eq.~\eqref{eq:lifetime} recovers the result of Esser et al.~\cite{esser_photoluminescence_2000} that has been previously derived for quantum well excitons.

\section{S10. Calculation of the dielectrically screened Coulomb potential}
We start by introducing the static nonlocal charge-density susceptibility $\chi(\br,\br')=\chi(\br,\br',\omega=0)$ that connects the material polarization 
$\boldsymbol{P}(\br)$ to the electric field $\boldsymbol{E}(\br)$ via
\begin{eqnarray}
    \boldsymbol{P}(\br) = \varepsilon_0 \int d^3\br'\chi(\br,\br')\boldsymbol{E}(\br')  \,.
\end{eqnarray}
The polarization is also related to the induced charge density $\rho_{\textrm{ind}}(\br)$ via
\begin{eqnarray}
\nabla_{\br} \cdot \boldsymbol{P}(\br)=-\rho_{\textrm{ind}}(\br)\, .
\end{eqnarray}
The total charge density can be expressed as a sum of free and induced charges:
\begin{eqnarray}
\rho(\br)=\rho_{\textrm{free}}(\br)+\rho_{\textrm{ind}}(\br)\, .
\end{eqnarray}
Using Gauss's law 
\begin{eqnarray}
\nabla_{\br} \cdot \boldsymbol{E}(\br)=-\, \frac{\rho(\br)}{\varepsilon_0}\, ,
\end{eqnarray}
we find
\begin{eqnarray}
\begin{split}
\varepsilon^{-1}_0\rho_{\textrm{free}}(\br)&=\varepsilon^{-1}_0\left(\rho(\br)-\rho_{\textrm{ind}}(\br)\right) \\ &=
\nabla_{\br} \cdot \boldsymbol{E}(\br) + \varepsilon^{-1}_0\nabla_{\br} \cdot \boldsymbol{P}(\br) \\
&=\nabla_{\br} \cdot \Big(\boldsymbol{E}(\br)+ \int d^3\br'\chi(\br,\br')\boldsymbol{E}(\br') \Big) \\
&=\nabla_{\br} \cdot \int d^3\br'\varepsilon_r(\br,\br')\boldsymbol{E}(\br')\, ,
\end{split}
\end{eqnarray}
where we introduced the dielectric function 
\begin{eqnarray}
    \varepsilon_r(\br,\br') = \delta(\br-\br') + \chi(\br,\br') \,.
\end{eqnarray}
With the effective electrostatic potential $\phi(\br)$ given by
\begin{eqnarray}
\boldsymbol{E}(\br)=-\nabla_{\br}\phi(\br)\, ,
\end{eqnarray}
a generalized form of Poisson's equation is derived:
\begin{eqnarray}
\nabla_{\br} \cdot\int d^3\br'\varepsilon_r(\br,\br') \nabla_{\br'}\phi(\br') = -\, \frac{\rho_{\textrm{free}}(\br)}{\varepsilon_0}\, .
\label{eq:poisson}
\end{eqnarray}
It yields the electrostatic potential $\phi$ for a given charge density $\rho_{\textrm{free}}$ in the presence of a dielectric function $\varepsilon_r$ describing nonlocal screening effects.
\begin{figure}[h]
\centering
\includegraphics[width=0.55\columnwidth]{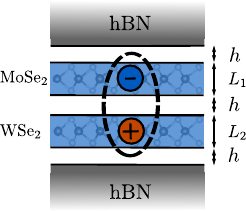}
\caption{Dielectric structure model for the interlayer Coulomb interaction.}
\label{fig:diel_struct}
\end{figure}

We consider a dielectric heterostructure as schematically shown in Fig.~\ref{fig:diel_struct}, where the hetero-bilayer extends in the x-y plane. Making use of the in-plane homogeneity of the continuum medium, Fourier transform with respect to in-plane coordinates yields a differential equation in the z coordinate only:
\begin{align}
\begin{split}
    &\int dz' \Big(\frac{\partial}{\partial z}\varepsilon(q,z,z')\frac{\partial}{\partial z'}\phi(q,z')-q^2 \varepsilon(q,z,z') \phi(q,z')\Big)\\ 
    &=\frac{\rho_{\textrm{free}}(q,z)}{\varepsilon_0}\, .
    \label{eq:poisson_1D}
    \end{split}
\end{align}
In the given geometry, where free charges can be located in different layers, nonlocality of the dielectric response, i.e. dependence of $\varepsilon$ on two different spatial coordinates, has to be considered. We simplify the spatial dependence of $\varepsilon$ by assuming that the hetero-bilayer can be approximated as an effective single layer with thickness $L=L_1 + L_2 + h $  and an averaged dielectric constant $\varepsilon_{\textrm{BL}}=(\varepsilon_{\textrm{MoSe}_2} L_1 + \varepsilon_{\textrm{WSe}_2} L_2 + h) / L $.
We further assume that the dielectric function of each layer is constant and local in z direction, i.e. $\varepsilon_r(q,z,z')= \varepsilon_r(q)\delta(z-z')$.
Then Poisson's equation can be solved analytically as shown in Ref.~\cite{florian_dielectric_2018} to obtain the macroscopic dielectric function $\varepsilon_q=\varepsilon_r(q)$.
An alternative approach to construct a dielectric function that is local in z-direction has been shown in Refs.~\cite{asriyan_optical_2019, semina_excitons_2019}.

\begin{figure}[h]
    \centering
    \includegraphics[width=0.55\columnwidth]{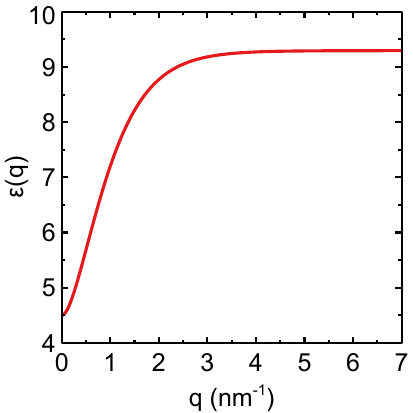}
    \caption{Calculated macroscopic dielectric function of the dielectric heterostructure shown in Fig.~\protect\ref{fig:diel_struct}. }
    \label{fig:eps_q}
\end{figure}

The Coulomb interaction between carriers located in different TMD layers is weaker than the intralayer Coulomb interaction due to the spatial separation of carriers in growth (z-)direction. To account for this effect the Coulomb matrix elements $V_q$ are modeled by using a formfactor $F_q$ in addition to the macroscopic dielectric function $\varepsilon_q$ according to
\begin{eqnarray}
    V_q = \frac{e^2}{2\varepsilon_0 q} \varepsilon^{-1}_q F_q\,.
\end{eqnarray}

The dielectric constants of the TMD materials are computed as geometric mean of the values given in~\cite{kylanpaa_binding_2015}, where also layer widths are provided. The dielectric constant of hBN is taken from~\cite{geick_normal_1966}. The interlayer distance $h = 0.3$ nm has been found to be an appropriate value in~\cite{florian_dielectric_2018}. 
The calculated dielectric function for dielectric heterostructure is shown in Fig.~\ref{fig:eps_q}. In the limit of small $q$ the dielectric function $\varepsilon_q$ resembles the Rytova-Keldysh limit~\cite{rytova1967the8248, keldysh_coulomb_1979}. Small deviations from the linear behavior arise from the interlayer distance of several \AA{} between TMD and the hBN layers~\cite{florian_dielectric_2018}. The form factor accounts for the confinement of carriers inside the atomically thin layers via the confinement functions $\xi(z)$
\begin{eqnarray}
    F_q = \int\mathrm{d}z\int\mathrm{d}z' \xi(z) \xi(z') e^{-iq|z-z'|}\xi(z')\xi(z)\,.
\end{eqnarray}
For the confinement functions, we assume eigenfunctions of the infinitely deep potential well with two nodes due to the mostly d-like character of electronic orbitals, which explicitly reads
\begin{eqnarray}
    F_q = \frac{4}{L_1 L_2}  \frac{ 324 \pi^4 e^{ - ( h + L_1 + L_2 ) q } 
     ( e^{ L_1 q } - 1 ) ( e^{ L_2 q } - 1 ) } { q^2 [ 36  \pi^2 + (L_1)^2 q^2 ][ 36 \pi^2 + (L_2)^2 q^2 ] }\,.
\end{eqnarray}

\bibliography{references}